\documentclass[review,12pt]{elsarticle}

\usepackage{amssymb}
 \usepackage{amsmath}

 \usepackage[text={6.4in,9.4in},centering]{geometry}

\newtheorem{prop}{Proposition} 
\newtheorem{lemma}[prop]{Lemma}
\newtheorem{theorem}[prop]{Theorem}

\newcommand{\pf}{\noindent{\em Proof. }}
\newcommand{\epf}{\hfill\hbox{\rule{6pt}{6pt}}\\}

\newcommand{\tw}{}

\begin{document}

\begin{frontmatter}



\title{UPGMA and the normalized equidistant minimum evolution problem}




\author[uea]{Vincent Moulton}
\ead{vincent.moulton@cmp.uea.ac.uk}

\author[as]{Andreas Spillner}
\ead{mail@andreas-spillner.de}

\author[uea]{Taoyang Wu \corref{cor1}}
\ead{taoyang.wu@gmail.com}

\address[uea]{
School of Computing Sciences,University of East Anglia, \\ Norwich,
NR4 7TJ, UK}

\address[as]{Department of Mathematics and Computer Science, University of Greifswald, Greifswald, Germany}

\cortext[cor1]{Corresponding author}

\begin{abstract}
UPGMA (Unweighted Pair Group Method with Arithmetic Mean) is a widely used clustering method. Here we
show that UPGMA is a greedy heuristic for  the normalized equidistant minimum evolution (NEME) problem, that is, finding a rooted
tree that minimizes the  minimum evolution score
relative to the dissimilarity matrix among all rooted trees with the same leaf-set in which all leaves have the same distance to the root. 
We prove that the NEME problem is NP-hard.  In addition, we 
present some heuristic and 
approximation algorithms for solving the NEME problem,
including  a polynomial time algorithm that yields
a binary, rooted tree whose NEME score
is within \(O(\log^2 n)\) of the optimum.
We expect that these results to eventually provide further insights into the behavior of the UPGMA algorithm.
\end{abstract}



\end{frontmatter}

\section{Introduction}
\label{section:introduction}

Clustering (i.e. subdividing a dataset into smaller subgroups or clusters) 
is a fundamental task in data analysis, and has a wide range of 
applications~(see, e.g.~\cite{d2005does}). An important 
family of clustering methods aim to produce a 
clustering of a dataset in which the clusters form a hierarchy
where the clusters nest within one another. Such hierarchies 
are typically represented by leaf-labeled tree structures known as  
dendrograms or rooted phylogenetic trees.
Introduced in 1958~\cite{sok-mic-58a}, average linkage analysis,
usually referred to as UPGMA (Unweighted Pair Group Method 
with Arithmetic Mean), is arguably the
most popular hierarchical clustering algorithm in
use to date, and remains  widely 
cited\footnote{According to Google Scholar, the method has been 
cited over 17,200 times during the period between 2011 and 2015.} and 
extremely popular~(see, e.g.~\cite{loewenstein2008efficient}). This is  
probably  because UPGMA is conceptually easy to understand
and fast in practice, an important consideration as big data sets
are becoming the norm in many areas. UPGMA is commonly used in 
phylogenetics and taxonomy to build
evolutionary trees~\cite[Chapter 11]{felsenstein2004inferring} 
as well as in related areas such as ecology~\cite{legendre2012numerical} and 
metagenomics~\cite{li2012ultrafast}. In addition, it is used 
as a general hierarchical clustering tool in bioinformatics and 
other areas including data mining and pattern 
recognition~\cite[Chapter 2]{romesburg2004cluster}.

UPGMA is a text-book algorithm that belongs to the family 
of agglomerative clustering methods that share the following 
common bottom-up scheme (cf. e.g. \cite[p.162]{felsenstein2004inferring}).
They take as input a dissimilarity $D$ on a set \(X\),
i.e. a real-valued, symmetric map on \( X \times X \)
which vanishes on the diagonal, and build
a collection of clusters or subsets of $X$ which 
correspond to a rooted tree with leaf-set $X$.
To do this, at each step two clusters with the minimum 
inter-cluster dissimilarity are combined to create a new cluster, 
starting with the collection
of clusters consisting of singleton subsets of $X$,
and finishing when the cluster $X$ is obtained.
Different formulations of the inter-cluster dissimilarity,  which 
specifies the dissimilarity of sets as a function of the 
dissimilarities observed on the members of the sets, lead to 
different heuristic criteria of the agglomerative methods. 
UPGMA, as the name average linkage analysis suggests, uses 
the mean dissimilarity across all pairs of elements that 
are contained within the two clusters.  
Formally, two clusters $A, B \subseteq X$ are selected for 
merging  at each iteration step of UPGMA if the average   
$$
\frac{1}{|A||B|} \sum_{a \in A,\, b \in B}  D(a,b)
$$
is minimized over all possible pairs of clusters. 
Since the arithmetic mean is used, UPGMA is often more stable 
than linkage methods in which only a subset of the 
elements within the clusters are used (e.g. the single-linkage method). 

UPGMA is commonly thought
of as a method that greedily constructs a rooted
phylogenetic tree that is closest to the input dissimilarity matrix in the 
least squares sense~\cite{dav-sul-12a}. However, it 
is not guaranteed to do so, although it often does quite well
in practice~\cite[p.162]{felsenstein2004inferring}. 
In \cite{gas-ste-06a} it was shown that
the related Neighbor-Joining \cite{sai-nei-87a} method for 
constructing unrooted phylogenetic trees from dissimilarity matrices 
can be thought of as a greedy heuristic that minimizes
the so-called \emph{balanced minimum evolution} score.
Here we shall observe that (see Section~\ref{section:basic:observations}), 
completely analogously, UPGMA is a 
greedy heuristic for computing a rooted
phylogenetic tree that minimizes the 
so-called \emph{minimum evolution} score \cite{rzh-nei-93a}
over all rooted phylogenetic trees on the same fixed leaf-set
in which all leaves have the same distance in the tree to the root.
We refer to this optimization problem as the
\emph{normalized equidistant minimum evolution} (NEME) problem, 
{\tw and expect that a better understanding of this problem will provide further insights into the behavior of the UPGMA algorithm}.

\subsection{Related work}

Theoretical properties of discrete optimization 
problems arising in the construction of
evolutionary trees have been studied for many years
(for some earlier work see, e.g. \cite{aga-baf-98a,day-87a,far-kan-95a}). 
Among these, the problems falling under the name of minimum evolution
alone form a quite diverse family (see, e.g. \cite{cat-09a}), in
which the so-called balanced minimum evolution 
problem~\cite{desper2004theoretical} is a particularly well-studied member.
For this problem it was recently shown in \cite{fio-jor-12a}
that for general \(n \times n\)-input dissimilarity matrices
there exists a constant \(c > 1\) such that no polynomial time algorithm
can achieve an approximation factor of \(c^n\) unless P equals NP.
We note that this hardness result 
does not rely on the often imposed restriction
(see, e.g.~\cite{bastkowski2015minimum,day-87a}) that the edge 
lengths of the constructed tree must be integers.
Moreover, in contrast to general input dissimilarity matrices, 
for inputs that are metrics (i.e. matrices
that also satisfy the triangle inequality) 
a polynomial 
time algorithm with an approximation factor of \(2\) is presented
in~\cite{fio-jor-12a}. Interestingly, the proof of this 
approximation factor uses the fact that
the balanced minimum evolution score
of an unrooted tree can be interpreted 
as being the average length of a spanning 
cycle compatible with the structure of the tree~\cite{sem-ste-04a}.

Another recent, related direction of work considers 
the algebraic structure of the space of rooted 
phylogenetic trees induced by the UPGMA 
method (see, e.g. \cite{dav-sul-12a,fah-hos-08a}).
This algebraic structure is tightly linked with the 
property of consistency of a tree construction method, that is,
those conditions under which the method is able to reconstruct
a tree that has been used to generate the input dissimilarity matrix (see, e.g. \cite{par-gas-12a}).
In the context of our work, we are particularly interested in 
the consistency of methods that perform a local search of the 
space of all rooted phylogenetic trees on a fixed set of leaves 
(see, e.g. \cite{desper2004theoretical}). 
Again, balanced minimum evolution is the variant of minimum
evolution for which some consistency results of 
this type are known \cite{bor-gas-09a,bor-mih-10a}.

\subsection{Our results and organization}

After presenting some preliminaries in the next section, in
Section~\ref{section:basic:observations} 
we begin by giving an explicit formula of the 
minimum evolution score of a rooted tree $T$ 
as a linear combination of the input dissimilarities.
This formula allows us to interpret
the minimum evolution score of  \(T\) in terms of the average length
of a minimum spanning tree compatible with the
set of clusters induced by \(T\). 

Using this observation, we explain how UPGMA can be
regarded as a greedy heuristic for the 
NEME problem. In addition, we show that
there are rooted phylogenetic trees with \(n\) leaves on which 
some input dissimilarity matrix has an optimal
least squares fit while the NEME score of that 
tree for the same dissimilarity matrix
is worse than the minimum NEME score by a factor 
in \(\Omega(n^2)\). This highlights the fact that the NEME 
problem and searching for trees
with minimum least squares fit are quite distinct
problems.

Next, in Section~\ref{section:searching:tree:space}, 
we explore solving the NEME problem by performing a local
search of the space of binary rooted phylogenetic trees 
using so-called rooted nearest neighbor interchanges 
as the moves in the local search. 
We show that this approach is consistent. More specifically,
for any input dissimilarity matrix that can be
perfectly represented by a unique binary rooted phylogenetic tree
with all leaves having the same distance from the root, we prove that
the local search will arrive at this tree after a finite
number of moves.

In Section~\ref{section:neme:np:hardness} we show that
the NEME problem is NP-hard even for 
\(n \times n\) input distance matrices
that satisfy the triangle inequality and only
take on \(O(\log n)\) different values.
In light of this fact, in Section~\ref{section:approximation} 
we consider some approximation algorithms
for solving the NEME problem. More specifically, we 
first show that the tree 
produced by UPGMA can have a score that
is worse than the minimum score by a factor in \(\Omega(n)\).
Then, for dissimilarity matrices that satisfy the triangle
inequality, we present a polynomial time algorithm that yields
a binary rooted phylogenetic tree whose NEME score
is within \(O(\log^2 n)\) of the optimum.
We conclude in Section~\ref{section:conclusion}
by mentioning two possible directions for future work.

\section{Preliminaries}
\label{section:definitions}

In this section we give a formal definition of the NEME problem
and introduce some of the notation and terminology that will
be used throughout this paper.

Let \(X\) be a finite non-empty set. A \emph{dissimilarity}
on \(X\) is a symmetric map 
\(D: X \times X \rightarrow \mathbb{R}\)
with \(D(x,x) = 0\) for all \(x \in X\).
In this paper,
a \emph{rooted phylogenetic tree} on \(X\) is
a rooted tree \(T=(V,E,\rho)\) with
(i) root \(\rho\) of degree~1, (ii) leaf set \(X\) and
(iii) every vertex not in \(X \cup \{\rho\}\)
having degree at least~3. Note that even though
we require the root to have degree~1, we do not consider
it a leaf of the tree. A 
\emph{normalized equidistant edge weighting} (NEEW) of a
rooted phylogenetic tree \(T=(V,E,\rho)\) on \(X\)
is a map \(\omega:E \rightarrow \mathbb{R}\)
such that the total weight of the edges
on the path from \(\rho\) to \(x\) is~0
for all \(x \in X\). More generally, for all \(u,v \in V\),
we denote the total weight
of the edges on the path between vertices \(u\) and \(v\)
by \(\ell_{(T,\omega)}(u,v)\). The \emph{height} \(h_{(T,\omega)}(v)\) of
any vertex \(v \in V\) is defined as \(\ell_{(T,\omega)}(v,x)\)
for any leaf \(x\) in the subtree of \(T\) rooted at \(v\).
The \emph{length} \(\ell_{\omega}(T)\)
of \(T\) under the edge weighting \(\omega\) is
\(\sum_{e \in E} \omega(e)\), that is, the total weight 
of all edges of~\(T\). 

Note that the length of any rooted phylogenetic tree 
\(T=(V,E,\rho)\) on \(X\) with a normalized equidistant edge 
weighting \(\omega\) can also be expressed as follows:
\begin{equation}
\label{equation:total:length:from:heights}
\ell_{\omega}(T) = \sum_{v \in (V-(X \cup \{\rho\}))} 
    (deg(v) - 2) \cdot h_{(T,\omega)}(v),
\end{equation}
where \(deg(v)\) denotes the degree of vertex \(v\) in \(T\).
Note that the restriction of \(\ell_{(T,\omega)}\) to
\(X \times X\) yields a dissimilarity \(D_{(T,\omega)}\) on \(X\).
Moreover, this dissimilarity is an \emph{ultrametric}, that is,  
\(D_{(T,\omega)}(x,z) \leq \max \{D_{(T,\omega)}(x,y),D_{(T,\omega)}(y,z)\}\)
holds for all \(x,y,z \in X\), if and only if
the edge weighting \(\omega\) assigns a non-negative
real number to each edge not adjacent to a vertex in
\(X \cup \{\rho\}\). We call any such edge weighting 
\emph{interior positive}.

Let \(D\) be a dissimilarity on \(X\) and \(T = (V,E,\rho)\)
a rooted phylogenetic tree on \(X\). For
any vertex \(v \in V\) let \(ch(v)\) denote the set of
\emph{children} of \(v\), that is, the set of vertices \(u\)
that are adjacent to \(v\) and for which \(v\) lies on the
path from \(u\) to \(\rho\). Moreover, we refer to \(v\) as the
\emph{parent} of the vertices \(u \in ch(v)\) and we
denote by \(C(v)\) the \emph{cluster} of elements in \(X\) induced by \(v\), that is,
the set of those leaves \(x\) of \(T\) for which the path
from \(x\) to \(\rho\) contains \(v\).
In \cite{far-69a} it is shown that, 
for any dissimilarity \(D\) and any rooted phylogenetic tree \(T\) on \(X\),
there is a unique normalized equidistant edge weighting
\(\omega = \omega_{(D,T)}\) with
\[\Delta(D,D_{(T,\omega)}) = \sum_{\{x,y\} \in \binom{X}{2}} \big ( D(x,y) - D_{(T,\omega)}(x,y) \big )^2\]
minimum, {\tw where $\binom{X}{2}$ denotes the set of 2-element subsets of $X
$.} More precisely, this edge weighting is the unique
solution of the system of linear equations
\begin{align}
\label{equation:optimal:edge:weights}
&\ell_{(T,\omega)}(v,y) \notag\\
&= \frac{1}{2 \sum_{\{u,u'\} \in \binom{ch(v)}{2}} |C(u)|\cdot|C(u')|}
\sum_{\{u,u'\} \in \binom{ch(v)}{2}} \sum_{\substack{x \in C(u) \\ x' \in C(u')}} D(x,x'),
\end{align}
for all \(v \in V - (X \cup \{\rho\})\),
and \(\ell_{(T,\omega)}(v,y) = 0\),
for all \(v \in X \cup \{\rho\}\), \(y \in C(v)\). 
Note that this is the analogue to
Vach's theorem for unrooted trees \cite{vac-89a}.
For later use, we put \(\Delta(D,D_{(T,\omega_{(D,T)})}) = \Delta(D,T)\).

Now, the normalized equidistant minimum evolution
score of a rooted
phylogenetic tree \(T\) on \(X\) with respect to a dissimilarity~\(D\)
on \(X\) is formally defined as
\[
\sigma_{D}(T) = \ell_{\omega_{(D,T)}}(T),
\]
that is, the length of \(T\) under the edge weighting
\(\omega_{(D,T)}\).
The NEME problem is to compute, for an input dissimilarity~\(D\) on \(X\), a rooted
phylogenetic tree on \(X\) with minimum NEME score. Formally, it can be stated as below. 

\medskip
\noindent
{\bf Problem } NEME Problem \\
{\bf Instance:} A distance matrix $D$ on a finite set $X$ and a number $p$. \\
{\bf Question:} Does there exist a rooted phylogenetic tree $T$ on $X$ such that $\sigma_{D}(T) \leq p$ holds.

\section{UPGMA and the NEME problem}
\label{section:basic:observations}

We begin this section by explaining how
the UPGMA algorithm can be reinterpreted as
a greedy approach to solving the NEME problem.
First note that it follows directly from
Equations~(\ref{equation:total:length:from:heights})
and~(\ref{equation:optimal:edge:weights})
that the NEME score of a rooted phylogenetic tree
\(T = (V,E,\rho)\) can be written as 
the following linear combination
of the given dissimilarity values:
\begin{align}
\label{formula:general:coefficients}
\sigma_{D}(T) &= \sum_{\{x,y\} \in \binom{X}{2}} \alpha_{\{x,y\}}^T \cdot D(x,y) 
\quad \text{with}\\
\alpha_{\{x,y\}}^T &= \frac{deg(lca(x,y))-2}{2 \sum_{\{u,u'\} \in \binom{ch(lca(x,y))}{2}} |C(u)|\cdot|C(u')|}\notag,
\end{align}
where \(lca(u,v)\) is the lowest common ancestor in \(T\)
for any two vertices \(u,v \in V\). In particular, in case \(T\)
is a \emph{binary} tree, that is, every vertex not in
\(X \cup \{\rho\}\) has degree precisely~3, 
we obtain, for any \(\{x,y\} \in \binom{X}{2}\) the coefficient
\begin{equation}
\label{formula:binary:coefficents}
\alpha_{\{x,y\}}^T = \frac{1}{2 \prod_{u \in ch(lca(x,y))} |C(u)|}.
\end{equation}
As an immediate consequence of (\ref{formula:general:coefficients}) we obtain
that the score \(\sigma_{D}(T)\) is linear in \(D\), that is,
when \(D\) can be written as \(D = \lambda_1 \cdot D_1 + \lambda_2 \cdot D_2\)
for non-negative real numbers \(\lambda_i\) and dissimilarities
\(D_i\), \(i \in \{1,2\}\), then
\begin{equation}
\label{equation:score:linear}
\sigma_{D}(T) = \lambda_1 \cdot \sigma_{D_1}(T) + \lambda_2 \cdot \sigma_{D_2}(T)
\end{equation}

To link the Formula~(\ref{formula:general:coefficients}) 
with the UPGMA algorithm, recall that this
algorithm constructs a rooted phylogenetic tree
by generating the list of clusters associated to
the vertices of this tree. It starts with the
singleton clusters associated to the leaves of
the tree. Then,
in each iteration of the algorithm we already have a partition 
\(\mathcal{C} = \{C_1,C_2,\dots,C_m\}\) of \(X\)
into \(m \geq 2\) clusters and a dissimilarity \(D\) on \(X\). 
UPGMA then selects a pair of two distinct
clusters \(A,B \in \mathcal{C}\) that minimizes
$$
\frac{1}{|A||B|} \sum_{a \in A, \, b \in B}  D(a,b)
$$
and returns the partition 
\((\mathcal{C} - \{A,B\}) \cup \{A \cup B\}\).
Now it is easy to see that UPGMA in each iteration
greedily pairs already selected clusters so
as to locally minimize the value added to
the score \(\sigma_{D}(T)\) for the rooted
phylogenetic tree~\(T\) produced by the method.

Interestingly, the coefficients in 
(\ref{formula:binary:coefficents}) suggest the
following alternative interpretation of the NEME score
of a binary rooted phylogenetic tree \(T=(V,E,\rho)\): Consider
the complete graph \(G\) with vertex set \(X\).
Each edge \(\{x,y\}\) of \(G\) is weighted with
the value \(D(x,y)\). We construct a random
subgraph of \(G\) as follows.
For each vertex \(v \in V - (X \cup \{\rho\})\)
select a random edge that has
precisely one end point
in each cluster associated with the two
children of \(v\). Let \(H\) denote
the resulting subgraph of \(G\). It is easy
to see that \(H\) is always a spanning
tree of \(G\).
Thus, in case \(T\) is binary, \(\sigma_{D}(T)\)
can be interpreted as half the average length
of a random spanning tree \(H\) of \(G\) that is compatible
with the clusters of \(T\).
Based on (\ref{formula:general:coefficients}),
this interpretation can be extended to the non-binary case
where, instead of a spanning tree, a random spanning
forest in \(G\) with \(|X|-1\) edges, some of which
selected more than once, arises. 

Next we present a technical lemma
summarizing some simple observations
about the NEME score that will be used later. 
Let \({\bf R}_X\) denote the
set of all rooted phylogenetic trees on \(X\). In
addition let \({\bf BR}_X\) denote the subset
of those trees in \({\bf R}_X\) that are binary.

\begin{lemma}
\label{lemma:basic:properties:NEME:score}
Let \(D\) be a non-negative dissimilarity on a finite set \(X\) with
\(|X| = n \geq 2\). Then, for all \(T \in {\bf R}_X\), we have:
\begin{itemize}
\item[{\rm (i)}]
\[
\sum_{\{x,y\} \in \binom{X}{2}} \alpha_{\{x,y\}}^T = \frac{1}{2} (n-1),
\]
\item[{\rm  (ii)}]
\[
\frac{2}{n^2} 
\leq \min \{ \alpha_{\{x,y\}}^T : \{x,y\} \in \binom{X}{2}\}
\leq \max \{ \alpha_{\{x,y\}}^T : \{x,y\} \in \binom{X}{2}\}
\leq \frac{1}{2}, 
\]
\item[{\rm (iii)}]
\[
\sigma_D(T) \leq \frac{n^2}{4} \min \{ \sigma_D(T') : T' \in {\bf R}_X\}.
\]
\end{itemize}
\end{lemma}

\pf
(i): We use induction on \(n\). For \(n=2\) the equality
clearly holds. Next assume \(n \geq 3\) and consider 
any \(T \in {\bf R}_X\). Let \(u\) be the single child of
\(\rho\). Put \(k=deg(u)-1\) and let \(v_1,v_2,\dots,v_k\)
denote the children of \(u\). Then we have, by induction,
\begin{align*}
\sum_{\{x,y\} \in \binom{X}{2}} \alpha_{\{x,y\}}^T 
&= \left [ \sum_{i=1}^{k} \sum_{\{x,y\} \in \binom{C(v_i)}{2}} \alpha_{\{x,y\}}^T \right ]
   +
   \left [ \sum_{\substack{\{x,y\} \in \binom{X}{2}\\u = lca(x,y)}} \alpha_{\{x,y\}}^T \right ]\\
&= \left [ \sum_{i=1}^{k} \frac{1}{2} (|C(v_i)| - 1) \right ] + \frac{k-1}{2}
= \frac{1}{2}(n-1),
\end{align*}
as required.

(ii): Consider any \(T \in {\bf R}_X\).
The inequality 
\(\max \{ \alpha_{\{x,y\}}^T : \{x,y\} \in \binom{X}{2}\} \leq \frac{1}{2}\)
follows immediately from the definition of the
coefficients \(\alpha_{\{x,y\}}^T\). And the inequality
\(\frac{2}{n^2} \leq \min \{ \alpha_{\{x,y\}}^T : \{x,y\} \in \binom{X}{2}\}\)
follows from the fact that, for any integer \(k \geq 2\), the function
\[ 
f:\mathbb{R}^k \rightarrow \mathbb{R}: 
(z_1,z_2,\dots,z_k) \mapsto \sum_{1 \leq i < j \leq k} z_iz_j
\]
attains its maximum among all non-negative 
\((z_1,z_2,\dots,z_k) \in \mathbb{R}^k\)
with \(\sum_{i=1}^k z_i = 1\) at \(z_1=z_2=\dots=z_k = \frac{1}{k}\).
Hence, for any \(\{x,y\} \in \binom{X}{2}\) with \(deg(lca(x,y)) - 1 = k\)
and \(|C(lca(x,y))| = m\), we have 
\(\alpha_{\{x,y\}}^T \geq \frac{k}{m^2}\).

(iii):
This is an immediate consequence of (ii):
\begin{align*}
\sigma_D(T) 
&\leq \sum_{\{x,y\} \in \binom{X}{2}} \frac{1}{2} D(x,y)\\
&= \frac{n^2}{4} \left [ \sum_{\{x,y\} \in \binom{X}{2}} \frac{2}{n^2} D(x,y) \right ]
\leq \frac{n^2}{4} \min \{ \sigma_D(T') : T' \in {\bf R}_X\}.
\end{align*}
\epf

We end this section presenting a family of dissimilarities \(D\)
for which the closest rooted equidistant
tree \(T\) in the least squares sense (i.e. the tree with
\(\Delta(D,T)\) minimum) has an NEME score that
is worse than the minimum NEME score by a quadratic factor.
This illustrates, as mentioned in the introduction, that
the NEME problem is quite different from the problem of
finding a closest rooted tree.

\begin{lemma}
\label{lemma:least:squares:lower:bound}
There exist dissimilarities \(D\) on a set \(X\)
with \(n\) elements for which there exists a 
rooted phylogenetic tree \(T\) on \(X\)
together with a normalized equidistant edge
weighting \(\omega\) with \(D(x,y) = D_{(T,\omega)}\)
for all \(x,y \in X\) but 
\[
\sigma_D(T) \geq \frac{n^2}{4} \min \{ \sigma_D(T') : T' \in {\bf R}_X\}
\]
\end{lemma}

\begin{figure}
\centering
\includegraphics[scale=1.0]{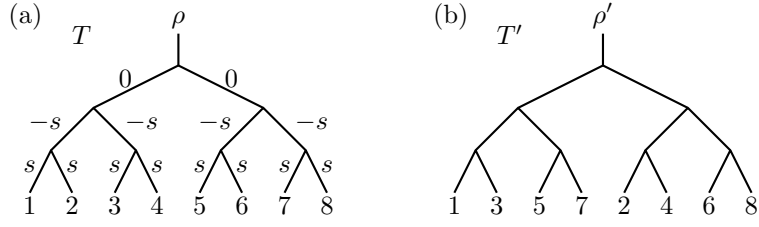}
\caption{Examples of rooted phylogenetic trees considered in the
         proof of Lemma~\ref{lemma:least:squares:lower:bound}.
         For the dissimilarity \(D=D_{(T,\omega)}\) induced by the rooted
         phylogenetic tree in (a) we have \(\sigma_D(T) =4s\).
         For the tree \(T'\) in (b) we obtain 
         \(\sigma_D(T') = \frac{s}{8}\).}
\label{figure:lower:bound:ls}
\end{figure}

\pf
Assume \(X=\{1,2,\dots,n\}\) with \(n = 2^k\) for some
integer \(k \geq 1\). Define the dissimilarity
\(D\) on \(X\) by putting \(D(i,i+1) = 2s\) for all
odd \(i \in X\), where \(s>0\) is a real number,
and \(D(x,y) = 0\) for all other \(x,y \in X\).
Note that in any rooted phylogenetic tree \(T\) on \(X\)
for which \(D=D_{(T,\omega)}\) holds with
\(\omega = \omega_{(D,T)}\) each pair \(\{i,i+1\}\),
\(i\) odd, must form a cherry 
(cf. Figure~\ref{figure:lower:bound:ls}(a)).
Moreover, for any such tree \(T\) we have
\(\sigma_D(T) = \frac{n}{2}s\). In contrast,
in any tree \(T'\) with minimum NEME score for \(D\)
the vertex \(lca(i,i+1)\) must be the single child of the root
for all odd \(i \in X\) (cf. Figure~\ref{figure:lower:bound:ls}(b)).
This implies \(\sigma_D(T') = \frac{2}{n}s\).
\epf

\section{Searching tree space for an optimal NEME tree}
\label{section:searching:tree:space}

In this section we shall first establish that performing a local search 
on the space \({\bf BR}_X\) for trees with minimum
NEME score is a {\em consistent} approach, that is,
if the input dissimilarity \(D\) can be represented by
a binary rooted phylogenetic tree with an interior positive
normalized equidistant edge weighting then this 
tree has minimum NEME score and, under some mild technical
conditions, the local search will
arrive at precisely this tree after a finite number of steps.

\begin{figure}
\centering
\includegraphics[scale=1.0]{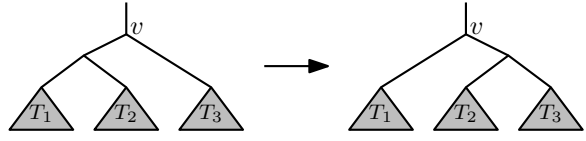}
\caption{A rooted nearest neighbor interchange prunes a
         subtree \(T_2\) from one child of \(v\), suppresses
         the resulting vertex of degree~2 and then grafts
         \(T_2\) onto the edge incident to the other child of~\(v\).}
\label{figure:example:rnni}
\end{figure}

Note that consistency is an important property and there are
general conditions known that imply consistency for
approaches that construct unrooted phylogenetic trees
(see e.g. \cite{par-gas-12a,wil-05a}).
We first show that for any \emph{generic} ultrametric, that is,
a dissimilarity \(D = D_{(T,\omega)}\) 
where  \(T=(V,E,\rho) \in {\bf BR}_X\)
and \(\omega\) is a normalized equidistant edge weighting
for \(T\) with \(\omega(e) > 0\) for all edges~\(e\)
not incident to a vertex in \(X \cup \{\rho\}\),
a local search in \({\bf BR}_X\) 
starting from any \(T' \in {\bf BR}_X\) using 
\emph{rooted nearest neighbor interchanges} (rNNI) will 
terminate in~\(T\). For unrooted trees an analogous result 
is established in \cite{bor-gas-09a}. Recall that an
rNNI modifies a rooted phylogenetic tree locally
around a vertex \(v\) as depicted in
Figure~\ref{figure:example:rnni}. In the following,
for any vertex \(v \neq \rho\) of a rooted phylogenetic tree
\(T=(V,E,\rho)\) on \(X\), the \emph{subtree} of \(T\) induced by \(v\)
consists of the parent \(p\) of \(v\) together with all the vertices \(u\) of \(T\) for which the
path from \(\rho\) to \(u\) contains \(v\). Note that such
a subtree can be viewed as a phylogenetic tree with root \(p\)
on the cluster \(C(v)\) of elements in \(X\) induced by \(v\).

In the following we will use the well known fact that,
for any generic ultrametric \(D\) on \(X\),
the binary rooted phylogenetic tree \(T \in {\bf BR}_X\)
with \(D = D_{(T,\omega)}\) for the edge weighting
\(\omega = \omega_{(D,T)}\) is unique \cite[Theorem 7.2.8]{sem-ste-03a}. We will say 
that \(T\) \emph{represents} \(D\), for short. 

\begin{lemma}
\label{lemma:nni:strictly decreases}
Let \(D\) be a generic ultrametric on \(X\) and 
\(T \in {\bf BR}_X\) the unique binary rooted
phylogenetic tree on \(X\) that represents~\(D\).
Then, for any \(T' \in ({\bf BR}_X - \{T\})\), there exists 
an rNNI that changes \(T'\) to \(T'' \in {\bf BR}_X\)
with \(\sigma_D(T') > \sigma_D(T'')\).
\end{lemma}

\begin{figure}
\centering
\includegraphics[scale=1.0]{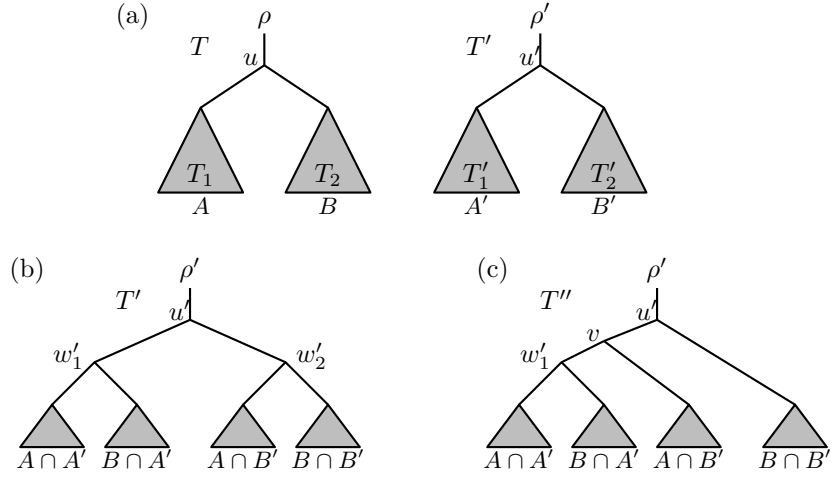}
\caption{(a)~The two trees \(T\) and \(T'\) considered in the
         proof of Lemma~\ref{lemma:nni:strictly decreases}.
         (b)~The detailed structure of \(T'\) in one of the
         cases considered in the proof.
         (c)~The tree~\(T''\) resulting from a suitable rNNI
         applied to~\(T'\).}
\label{figure:basic:situation:consistency}
\end{figure}

\pf
We use induction on \(n = |X|\). The statement in the lemma
clearly holds for \(n \in \{1,2\}\) in view of the fact
that \(|{\bf BR}_X|=1\). So assume \(n \geq 3\) and
consider any \(T' \in ({\bf BR}_X - \{T\})\).
The situation is depicted in 
Figure~\ref{figure:basic:situation:consistency}(a).
Let \(A\) and \(B\), respectively, denote the set of leaves
in the rooted subtrees \(T_1\) and \(T_2\) of \(T\).
Similarly, let \(A'\) and \(B'\), respectively, denote the
set of leaves in the rooted subtrees \(T_1'\) and \(T_2'\) of \(T'\).
Note that the restriction \(D_{|Y}\) of \(D\) to any non-empty subset 
\(Y \subseteq X\) is again a generic ultrametric.

First consider the case that \(T_1'\) does not 
represent \(D_{|A'}\). Then, by induction, there
exists an rNNI in \(T_1'\) that results in a rooted
phylogenetic tree \(T_1''\) on \(A'\) with strictly
smaller NEME score. Hence, applying the same rNNI 
to \(T'\) yields a tree \(T''\) with 
\(\sigma_D(T') > \sigma_D(T'')\). The case that
\(T_2'\) does not represent \(D_{|B'}\) is completely
analogous.

It remains to consider the case that
\(T_1'\) and \(T_2'\) represent
\(D_{|A'}\) and \(D_{|B'}\), respectively.
Then, \(A=A'\) and \(B=B'\) immediately implies 
\(T_1=T_1'\) and \(T_2 = T_2'\) and, thus, \(T=T'\).
Otherwise, there exists at least one \(\{x,y\} \in \binom{X}{2}\)
with \(\{x,y\} \subseteq A\) or \(\{x,y\} \subseteq B\)
but \(x \in A'\) and \(y \in B'\).
Thus, swapping the roles of either \(A\) and \(B\) or \(A'\) and \(B'\),
we can assume without loss of generality that
the sets \(A \cap A'\), \(A \cap B'\) and
\(B \cap B'\) are non-empty. The structure of
the tree \(T'\) is 
depicted in Figure~\ref{figure:basic:situation:consistency}(b).
Put \(\omega' = \omega_{(D,T')}\).

First assume that \(B \cap A' \neq \emptyset\).
Put \(n_1 = |A \cap A'|\), \(n_2 = |B \cap A'|\),
\(n_3 = |A \cap B'|\) and \(n_4 = |B \cap B'|\).
Without loss of generality we assume \(n_1 + n_2 > n_4\).
We perform an rNNI pruning and regrafting the
subtree with leaf set \(A \cap B'\) to obtain
the tree \(T''\) depicted in 
Figure~\ref{figure:basic:situation:consistency}(c).
Put \(\omega'' = \omega_{(D,T'')}\).
To show \(\sigma_D(T') > \sigma_D(T'')\)
it suffices to show
\(h_{(T',\omega')}(u') + h_{(T',\omega')}(w_2')
> h_{(T'',\omega'')}(u') + h_{(T'',\omega'')}(v)\).
To establish the latter inequality, recall that \(T\) represents \(D\) and, therefore,
we can assume that \(D\) has been scaled
so that \(D(a,b) = 1\) for all \(a \in A\), \(b \in B\).
We put 
\[
\delta_{A} = \sum_{\substack{a' \in (A \cap A')\\b' \in (A \cap B')}} D(a',b') 
\quad \text{and} \quad
\delta_{B} = \sum_{\substack{a' \in (B \cap A')\\b' \in (B \cap B')}} D(a',b').
\]
Note that all distances that contribute to \(\delta_{A}\) and \(\delta_{B}\)
are strictly less than~\(1\), implying \(\delta_{A} < n_1n_3\)
and \(\delta_{B} < n_2n_4\). Using this notation, we obtain
\begin{align*}
&h_{(T',\omega')}(u') + h_{(T',\omega')}(w_2') 
   - h_{(T'',\omega'')}(u') - h_{(T'',\omega'')}(v)\\
&= \frac{1}{2} + \frac{n_1n_4+n_2n_3}{2(n_1+n_2)(n_3+n_4)}
   - \frac{n_2}{2(n_1+n_2)} - \frac{n_1+n_3}{2(n_1+n_2+n_3)}\\
&\quad + \frac{1}{2} \left [ \frac{1}{(n_1+n_2)(n_3+n_4)}
                        -\frac{1}{(n_1+n_2)n_3} \right ] \cdot \delta_{A}\\
&\quad + \frac{1}{2} \left [ \frac{1}{(n_1+n_2)(n_3+n_4)} 
                        -\frac{1}{(n_1+n_2+n_3)n_4} \right ] \cdot \delta_{B}\\
&= g(\delta_{A},\delta_{B}).
\end{align*}
Note that \(g(\delta_{A},\delta_{B})\) is a linear function
in \(\delta_{A}\) and \(\delta_{B}\) and that the coefficient
of \(\delta_{A}\) is negative. Moreover, the assumption
\(n_1+n_2 > n_4\) implies that the coefficient of \(\delta_{B}\)
is negative too. Thus, using the fact that \(\delta_{A} < n_1n_3\)
and \(\delta_{B} < n_2n_4\), we have
\[g(\delta_{A},\delta_{B}) > g(n_1n_3,n_2n_4) = 0,\]
from which \(\sigma_D(T') > \sigma_D(T'')\) follows, as required.

It remains to consider the case that \(B \cap A' = \emptyset\),
that is, \(n_2 = 0\).
We apply the same rNNI to \(T'\) as in the previous case and,
using similar calculations, we obtain
\begin{align*}
&\sigma_D(T') - \sigma_D(T'')
= \frac{n_4}{2(n_3+n_4)} + \frac{1}{2} 
  \left [ \frac{1}{n_1(n_3+n_4)} - \frac{1}{n_1n_3} \right ] \cdot \delta_{A}
 > 0,
\end{align*}
using again \(\delta_{A} < n_1n_3\).
\epf

In the following main result of this section
we note that even for the non-generic
case a weak form of consistency holds.

\begin{theorem}
\label{theorem:consistency:NEME:binary:trees}
Let \(T=(V,E,\rho) \in {\bf BR}_X\) and \(\omega\) an
interior-positive normalized equidistant
edge weighting for \(T\). Put \(D = D_{(T,\omega)}\).
Then we have
\[
\sigma_D(T) = \min \{ \sigma_D(T') : T' \in {\bf BR}_X\}.
\]
If \(D\) is generic, then \(T\) is the unique tree
in \({\bf BR}_X\) minimizing the NEME score for \(D\)
and a local search using rooted nearest neighbor interchanges
starting from any tree in \({\bf BR}_X\) 
will arrive at \(T\) after a finite number of steps.
\end{theorem}

\pf
For generic \(D\), the theorem is an immediate consequence 
of Lemma~\ref{lemma:nni:strictly decreases}. So, assume that
\(D\) is not generic and, for a contradiction, 
that there exists some \(T' \in {\bf BR}_X\)
with \(\sigma_D(T') < \sigma_D(T)\). For any real number 
\(\varepsilon > 0\), define the NEEW \(\omega_{\varepsilon}\)
of \(T\) by putting \(\omega_{\varepsilon}(e)=\omega(e)+\varepsilon\)
for all edges \(e\) of \(T\) not adjacent to a vertex in \(X \cup \{\rho\}\) and
\(\omega_{\varepsilon}(e)=\omega(e)\) for all other edges \(e\) of \(T\).
Put \(D_{\varepsilon} = D_{(T,\omega_{\varepsilon})}\) and note that, by
construction, \(D_{\varepsilon}\) is a generic ultrametric that is
represented by \(T\). As a consequence, 
\(\sigma_{D_{\varepsilon}}(T) < \sigma_{D_{\varepsilon}}(T')\) must hold.
But this contradicts \(\sigma_D(T') < \sigma_D(T)\) 
in view of the fact that, as \(\varepsilon\) tends to \(0\),
\(\sigma_{D_{\varepsilon}}(T)\) tends to \(\sigma_D(T)\) while
\(\sigma_{D_{\varepsilon}}(T')\) tends to \(\sigma_D(T')\).
\epf

The result in Theorem~\ref{theorem:consistency:NEME:binary:trees}
immediately raises the question whether the 
NEME score for any input distance matrix is minimized by some
binary rooted phylogenetic tree. It is known \cite{eic-hug-08} that
for unrooted phylogenetic trees the balanced minimum evolution score
is indeed always minimized for some unrooted binary phylogenetic tree.
We end this section establishing that, in contrast, the answer to the above
question for the NEME score is no.

\begin{lemma}
\label{lemma:non:binary:trees:better}
There exist dissimilarities \(D\) with
\[
\min \{ \sigma_D(T) : T \in {\bf R}_X\} < 
\min \{ \sigma_D(T) : T \in {\bf BR}_X\}.
\]
\end{lemma}

\begin{figure}
\centering
\includegraphics[scale=1.0]{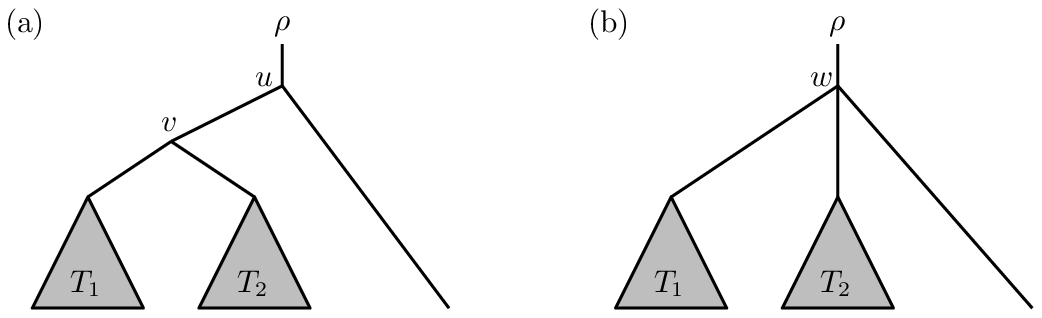}
\caption{The construction referred to in the proof of
         Lemma~\ref{lemma:non:binary:trees:better}.}
\label{figure:non:binary:better}
\end{figure}

\pf
Consider a binary rooted phylogenetic tree \(T\)
whose structure is as depicted in 
Figure~\ref{figure:non:binary:better}(a).
It consists of two rooted binary subtrees \(T_1\) and
\(T_2\), each having \(m \geq 3\) leaves. In addition,
there is a single leaf adjacent to vertex \(u\).
Let \(\omega\) be an interior positive
normalized equidistant edge weighting for \(T\)
such that \(h_{(T,\omega)}(v) = 1\) and 
\(h_{(T,\omega)}(u) = s\) for some \(s > 1\).
Put \(D = D_{(T,\omega)}\).

Next, consider the non-binary  rooted phylogenetic
tree \(T'\) depicted in Figure~\ref{figure:non:binary:better}(b).
It is constructed from \(T\) by contracting the
edge between \(u\) and \(v\) into the
vertex \(w\). Put \(\omega' = \omega_{(D,T')}\).
To show that \(\sigma_D(T) > \sigma_D(T')\),
it suffices, by Equation~(\ref{equation:total:length:from:heights}),
to show that
\begin{align*}
h_{(T,\omega)}(u) + h_{(T,\omega)}(v) 
&= s + 1
> 2 h_{(T',\omega')}(w) = \frac{2m^2+4ms}{m^2+2m},
\end{align*}
which can easily be checked to be the case,
in view of \(m \geq 3\), for any \(s > 1\). Hence, by 
Theorem~\ref{theorem:consistency:NEME:binary:trees},
we have
\(\min \{ \sigma_D(T') : T' \in {\bf R}_X\} < 
\sigma_D(T) =
\min \{ \sigma_D(T') : T' \in {\bf BR}_X\}\).
\epf

\section{The NEME problem is NP-hard}
\label{section:neme:np:hardness}

To establish NP-hardness of the NEME problem, we use
a reduction from the well-known NP-hard graph coloring problem 
(see e.g.~\cite{gar-joh-79a}). More specifically, we consider
the following variant of this problem:
\smallskip\\
\noindent
\textsc{Input}: A graph \(G=(V,E)\) with \(|V| = 4n\).\\
\noindent
\textsc{Question}: Can \(V\) be partitioned into 
4~subsets \(V_1,V_2,V_3,V_4\) with 
\(|V_1| = |V_2| = |V_3| = |V_4| = n\) such that
no edge \(e \in E\) has both endpoints in the same
set \(V_i\) for some \(i \in \{1,2,3,4\}\)? We call any
such partition a \emph{4-coloring} of \(G\).
\smallskip

Note that the additional constraint that the sets in 
a 4-coloring are all of the same size is merely added to simplify
the description of the reduction. It preserves the
NP-hardness of the graph coloring problem in view of the
fact that adding isolated vertices to any graph \(G\) does not
change the minimum number of colors that suffice to
color~\(G\). We first present a technical lemma that will be used
in the construction below.

\begin{lemma}
\label{lemma:distance:shrinking}
Let \(X\) be a set with \(2(m+k)\) elements, \(m \geq 1\), \(k \geq 1\),
that is partitioned into the sets \(A\), \(B\) and \(C\) with
\(|A|=|B|=m\) and \(|C| = 2k\). In addition, let \(s > 0\) be a real
number and \(D\) a dissimilarity on \(X\) with
\(D(x,y) = s\) for all \(x \in A\), \(y \in B\) and
\(D(x,y) \leq \frac{s}{3(m+k)^5}\) for all other \(x,y \in X\).
Then any binary rooted phylogenetic tree \(T = (V,E,\rho)\) on \(X\)
with \(\sigma_D(T) = \min \{ \sigma_D(T) : T \in {\bf BR}_X\}\)
must contain two distinct vertices \(v,w \in V\) with
(i) \(C(v) \cap C(w) = \emptyset\),
(ii) \(|C(v)| = |C(w)| = (m+k)\),
(iii) \(A \subseteq C(v)\) and \(B \subseteq C(w)\).
\end{lemma}

\pf
First consider a binary rooted phylogenetic tree \(T\)
on \(X\) that contains vertices \(v\) and \(w\) with
properties (i)-(iii). Note that this implies that
\(v\) and \(w\) have the same parent \(u\) and that
\(u\) is the single child of the root \(\rho\).
Moreover, for all \(a \in A\), \(b \in B\), we have
\(\alpha^T_{\{a,b\}} = \frac{1}{2(m+k)^2}\) and,
by Lemma~\ref{lemma:basic:properties:NEME:score}(ii),
this is the smallest possible value for a rooted phylogenetic
tree with \(2(m+k)\) leaves.

Next consider any binary rooted phylogenetic tree \(T' = (V',E',\rho')\)
on \(X\) that does not contain two vertices \(v\) and \(w\)
satisfying properties (i)-(iii). Let \(u'\) denote the single
child of \(\rho'\) and consider the two children \(v'\) and \(w'\)
of \(u'\). In particular, \(v'\) and \(w'\) must violate
at least one of the properties (i)-(iii). By construction
we have \(C(v') \cap C(w') = \emptyset\). Hence, one of the
properties (ii) or (iii) must be violated.

First consider the case that (iii) is violated. This implies,
without loss of generality, that there exist \(a \in A\) and
\(b \in B\) with \(\{a,b\} \subseteq C(v')\). In view of
\(|C(v')| \leq 2(m+k)-1\) and 
Lemma~\ref{lemma:basic:properties:NEME:score}(ii) we have
\[
\alpha^{T'}_{\{a,b\}} 
\geq \frac{2}{(2(m+k)-1)^2}
= \frac{1}{2(m+k)^2 - 2(m+k) + 1/2}
\]
Next consider the case that property (iii) is satisfied
but (ii) is violated for \(v'\) and \(w'\).
Then, for any \(a \in A\) and any \(b \in B\), we have
\[
\alpha^{T'}_{\{a,b\}} 
\geq \frac{1}{2(m+k+1)(m+k-1)}
= \frac{1}{2(m+k)^2 - 2}
\]

Noting that \(2(m+k)^2 - 2 \geq 2(m+k)^2 - 2(m+k) + 1/2\),
we calculate a lower bound on the difference between the
coefficients in \(T\) and \(T'\):
\[
\frac{1}{2(m+k)^2 - 2} - \frac{1}{2(m+k)^2}
\geq \frac{1}{2(m+k)^4}.
\]
This implies, using Lemma~\ref{lemma:basic:properties:NEME:score}(i)
to obtain the upper bound in the second line below:
\begin{align*}
\sigma_D(T) 
&= \sum_{\{x,y\} \in \binom{X}{2}} \alpha^T_{\{x,y\}} D(x,y)\\
&\leq \left [ \sum_{a \in A, \, b \in B} \alpha^T_{\{a,b\}} D(a,b) \right ] + 
      \frac{s(m+k)}{3(m+k)^5}\\
&< \left [ \sum_{a \in A, \, b \in B} \alpha^T_{\{a,b\}} s \right ] +
      \frac{s}{2(m+k)^4}\\
&\leq \sum_{a \in A,\, b \in B} \alpha^{T'}_{\{a,b\}} s 
\leq \sum_{\{x,y\} \in \binom{X}{2}} \alpha^{T'}_{\{x,y\}} D(x,y)
= \sigma_D(T')
\end{align*}
Hence, \(T'\) cannot be an optimal tree for \(D\).
\epf

Next we describe how to construct for a given
graph \(G =(V(G),E(G))\) with \(|V(G)| = 4n\)
a suitable dissimilarity \(D=D(G)\). 
First construct a set \(X\) that is the disjoint union
of \(V(G)\), \(Y\) and \(W\) with \(|Y| = 2^k\) and
\(|W| = (2^k - 4)n\) where \(k = \lfloor \log_2(n+1) + 4 \rfloor\).
Note that 
\[
|X| = 2^k(n+1) \leq 2^{\log_2(n+1) + 4}(n+1) = 16(n+1)^2.
\]
Put \(m_i = 2^{k-i}(n+1)\), \(i \in \{0,1,\dots,k\}\).
In addition, put \(s_1 = 1\) and, for \(i \in \{1,2,\dots,k\}\),
\(s_{i+1} = \frac{s_i}{3m_i^5}\). Moreover, put
\(s^* = \frac{s_{k+1}}{2^{k+2}(n+1)^3}\). The values
\(s_1 > s_2 >\dots > s_{k+1} > s^* > 0\) will be the
possible distances between elements in \(X\).

Now, recursively partition the set \(Y\) so as to force
a fully balanced binary tree as a backbone structure.
More precisely, put \(Y_{0,0} = Y\) and define, 
for all \(i \in \{0,1,\dots,k-1\}\)
and all \(j \in \{0,1,\dots,2^{i}-1\}\),
sets \(Y_{i+1,2j}\) and \(Y_{i+1,2j+1}\) so that
\(Y_{i+1,2j} \cap Y_{i+1,2j+1} = \emptyset\), 
\(Y_{i+1,2j} \cup Y_{i+1,2j+1} = Y_{i,j}\) and
\(|Y_{i+1,2j}| = |Y_{i+1,2j+1}|\) hold.
Select an element \(y_l^* \in Y_{2,l}\) for
each \(l \in \{0,1,2,3\}\).

\begin{figure}
\centering
\includegraphics[scale=1.0]{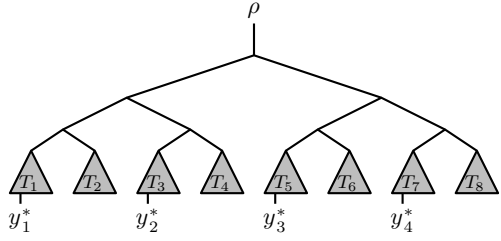}
\caption{The structure of the backbone tree for \(k=3\).}
\label{figure:backbone:structure}
\end{figure}

Next we construct the dissimilarity \(D = D(G)\) on \(X\):
\begin{itemize}
\item[(a)]
For all \(w \in W\) and all \(x \in X\) we put \(D(w,x) = 0\).
The elements in \(W\) are just used to fill subtrees so that
we get a fully balanced backbone tree.
\item[(b)]
For all \(y,y' \in Y\), \(y \neq y'\), we put \(D(y,y') = s_{i+1}\)
where \(i\) is the largest index in \(\{0,1,\dots,k-1\}\)
with \(\{y,y'\} \subseteq Y_{i,j}\) for some 
\(j \in \{0,1,\dots,2^i-1\}\). The distances between the
elements in \(Y\) force a fully balanced backbone tree
by Lemma~\ref{lemma:distance:shrinking}
(cf. Figure~\ref{figure:backbone:structure}).
\item[(c)]
For all \(v \in V(G)\) and all \(y \in (Y - \{y_1^*,y_2^*,y_3^*,y_4^*\})\) 
we put \(D(v,y) = s_{k+1}\). And for all
\(v \in V(G)\) and all \(y \in \{y_1^*,y_2^*,y_3^*,y_4^*\}\)
we put \(D(v,y) = 0\).
This will force that a subset of \(n\) vertices 
of \(G\) is grouped together with each \(y_l^*\), \(1 \leq l \leq 4\),
in the same subtree.
\item[(d)]
For all \(v,v' \in V(G)\), \(v \neq v'\),
we put \(D(v,v') = s^*\) if \(\{v,v'\} \in E(G)\) and
\(D(v,v') = 0\) otherwise. These distances capture the
structure of \(G\) and are so small that they do not
interfere with forming the fully balanced backbone tree.
\end{itemize}

\begin{lemma}
\label{lemma:reduction:works}
Let \(G\) be a graph with \(4n\) vertices
and let \(D=D(G)\) be the dissimilarity on \(X\)
constructed above. Then
\(G\) has a 4-coloring if and only if there
exists a binary rooted phylogenetic tree \(T\)
on \(X\) with \(\sigma_{D}(T)\) not larger than
\[
\left [
2^{k-2} \sum_{i=1}^k \frac{s_i}{m_i^2}
\right ]
+ 
s_{k+1} \cdot 4n \cdot
\left [
\left ( \sum_{i=0}^{k-3} \frac{2^{i}}{2m_{k-i}^2} \right )
+ \frac{2^{k-2}-1}{2m_2^2} + \frac{2^{k-1} - 2}{2m_1^2}
\right ]
+ \frac{s^*}{2^{2k-7}}
\]
\end{lemma}

\pf
First note that, by the construction of
the distances \(s_1,s_2,\dots,s_{k+1},s^*\) and
Lemma~\ref{lemma:distance:shrinking}, the upper part of any
optimal binary rooted phylogenetic tree \(T\)
on \(X\) must be a fully balanced binary tree.
This upper part has \(k+1\) levels.
Level \(i \in \{0,1,\dots,k\}\) consists
of \(2^i\) subtrees, each of which containing
precisely one of the sets \(Y_{i,j}\), \(j \in \{0,1,\dots,2^{i}-1\}\)
in its set of leaves. In particular,
the lowest level consists of \(2^k\) subtrees
\(T_1,T_2,\dots,T_{2^k}\) and the leaf set of each of
these subtrees consists of precisely one element from \(Y\)
and \(n\) elements from \(V(G) \cup W\).
Thus, in view of \(y^*_l \in Y_{2,l}\), \(l \in \{0,1,2,3\}\),
we can choose the numbering of the subtrees so that
\(T_{l2^{k-2}+1}\) is the subtree that contains 
leaf \(y^*_l\) (cf. Figure~\ref{figure:backbone:structure}).

Now consider any vertex \(v \in V(G)\). First assume
that \(v\) is a leaf in one of the subtrees \(T_{l2^{k-2}+1}\), 
\(l \in \{0,1,2,3\}\).
Then, because the backbone tree
is fully balanced, the contribution to \(\sigma_D(T)\)
of the distances from \(v\) to the elements in \(Y\) is
\[
s_{k+1} \cdot
\left [
\left ( \sum_{i=0}^{k-3} \frac{2^{i}}{2m_{k-i}^2} \right )
+ \frac{2^{k-2}-1}{2m_2^2} + \frac{2^{k-1} - 2}{2m_1^2}
\right ].
\]
Next assume that \(v\) is a leaf in some subtree
\(T_q\), \(q \neq l2^{k-2}+1\) for all \(l \in \{0,1,2,3\}\). 
Then one summand of the
form \(\frac{s_{k+1}}{2m_{k-i}^2}\), \(i \in \{0,1,\dots,k-3\}\),
is replaced by a summand that contributes, 
by Equation~(\ref{formula:general:coefficients}), at least
\(\frac{1}{2(m_k/2)^2} \cdot s_{k+1} = \frac{2s_{k+1}}{m_k^2} = \frac{2s_{k+1}}{(n+1)^2}\).
Thus, the increase in the contribution is at least
\[
\left [ \frac{2}{m_k^2} - \frac{1}{2m_k^2} \right ] \cdot s_{k+1}
= \frac{3s_{k+1}}{2(n+1)^2} > 2^{k-1}(n+1)s^* 
\geq \sum_{\{x,y\} \in \binom{X}{2}} \alpha^{T}_{\{x,y\}}s^*,
\]
that is, using Lemma~\ref{lemma:basic:properties:NEME:score}(i)
to obtain the last inequality,
it is strictly larger than the total contribution
of all distances that equal~\(s^*\).
Hence, the contribution of the distances \(s_{k+1}\)
to the score is minimized if and only if each vertex
\(v \in V(G)\) is a leaf in one of the subtrees \(T_l\),
\(1 \leq l \leq 4\), inducing a partition of
\(V(G)\) into 4 subsets \(V_1,V_2,V_3,V_4\) each of
size \(n\).

Summarizing the contribution
of the distances \(s_1,s_2,\dots,s_{k+1}\) to the score
of the tree, we obtain:
\begin{align*}
\lambda(n,k) =
\left [
2^{k-2} \sum_{i=1}^k \frac{s_i}{m_i^2}
\right ]
+ 
s_{k+1} \cdot 4n \cdot
\left [
\left ( \sum_{i=0}^{k-3} \frac{2^{i}}{2m_{k-i}^2} \right )
+ \frac{2^{k-2}-1}{2m_2^2} + \frac{2^{k-1} - 2}{2m_1^2}
\right ]
\end{align*}
Note that this contribution does \emph{not}
depend on the structure of the graph~\(G\).

It remains to calculate the contribution of the distances
that equal \(s^*\) to the score of the tree.
Note that \(16(n+1)^2\) is a trivial upper bound
on the number of edges in~\(G\). Thus,
if the partition \(V_1,V_2,V_3,V_4\)
induced by the tree is a 4-coloring of \(G\), then
the total contribution to the score of the tree is at most
\[
\frac{16(n+1)^2s^*}{2m_2^2} = \frac{16s^*}{2^{2k-3}} = \frac{s^*}{2^{2k-7}}.
\]
In contrast, a single edge with both endpoints in one
of the sets \(V_l\), \(1 \leq l \leq 4\),
contributes, by Equation~(\ref{formula:general:coefficients}),
at least \(\frac{1}{2((n+1)/2)^2}s^* = \frac{2s^*}{(n+1)^2}\).
Hence, noting that \(k = \lfloor \log_2(n+1) + 4 \rfloor\)
implies \(\frac{2}{(n+1)^2} > \frac{1}{2^{2k-7}}\),
we obtain that \(G\) has a 4-coloring if and only
if there exists a binary rooted phylogenetic tree \(T\)
on \(X\) with
\[
\sigma_D(T) \leq \lambda(n,k) + \frac{s^*}{2^{2k-7}}.
\]
\epf

Note that the dissimilarity \(D=D(G)\) constructed
above need not satisfy the triangle inequality.
However, putting \(D'(x,x') = D(x,x') + 1\) for
all \(x,x' \in X\), \(x \neq x'\), and 
\(D'(x,x) = 0\) for all \(x \in X\), 
we obtain a dissimilarity \(D'\) on \(X\)
that satisfies the triangle inequality.
Moreover, by Lemma~\ref{lemma:basic:properties:NEME:score}(i),
for every binary rooted phylogenetic tree
\(T\) on \(X\), we have
\(\sigma_{D'}(T) = \sigma_D(T) + \frac{1}{2}(|X| - 1)\),
that is, a tree \(T\) is optimal for~\(D\) if and only
if \(T\) is optimal for~\(D'\). Thus we have the main
result of this section:

\begin{theorem}
\label{theorem:summary:NEME:hardness}
Computing a binary rooted phylogenetic tree
with minimum NEME score for a dissimilarity~\(D'\)
on a set \(X\) is NP-hard even if \(D'\)
satisfies the triangle inequality and
\(D'\) takes on only \(O(\log_2(|X|))\)
different values.
\end{theorem}

\section{Approximating the minimum NEME score}
\label{section:approximation}

Note that Lemma~\ref{lemma:basic:properties:NEME:score}(iii)
states that \emph{any} tree in \({\bf R}_X\) approximates
the minimum NEME score over all trees in \({\bf R}_X\) up
to a factor that is in \(O(n^2)\), \(n=|X|\). It is not hard to
see that this bound is asymptotically tight and in this section we
explore ways to obtain better approximation guaranties. 

We first look at the approximation gaurantees that can be achieved
with existing algorithms. We start with UPGMA and
establish a lower bound of \(\Omega(n)\) on the 
approximation guaranty achieved by it.

\begin{lemma}
\label{lemma:lower:bound:UPGMA}
For every non-empty finite set \(X\) 
with \(n \geq 3\) elements there exists a
dissimilarity \(D\) on \(X\) such that
\[
\sigma_D(T) \geq \frac{n}{4} \min \{ \sigma_D(T') : T' \in {\bf R}_X\}
\]
holds for the tree \(T\) produced by UPGMA.
\end{lemma}

\begin{figure}
\centering
\includegraphics[scale=1.0]{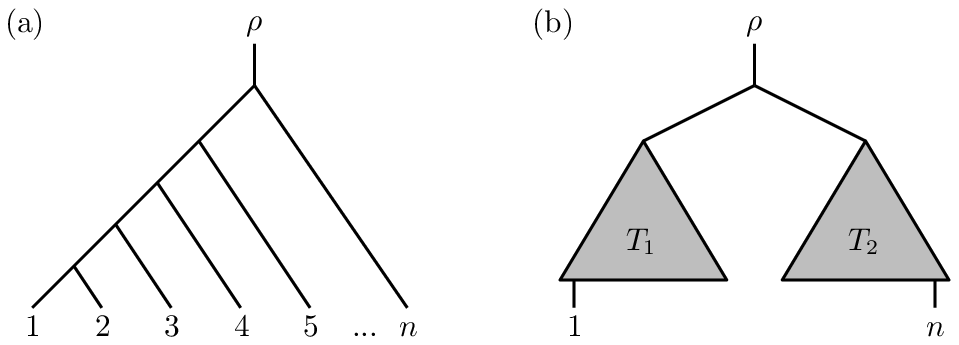}
\caption{Two rooted phylogenetic trees used in the proof
         of Lemma~\ref{lemma:lower:bound:UPGMA}.}
\label{figure:lower:bound:upgma}
\end{figure}

\pf
Let \(X=\{1,2,\dots,n\}\) and consider the ultrametric
\(D'\) on \(X\) defined by putting, for all \(1 \leq i < j \leq n\),
\(D'(i,j) = D'(j,i) = 1 + \frac{(j-2)}{n-2}\). 
Note that the unique rooted phylogenetic tree \(T\)
on \(X\) with \(D_{(T,\omega)} = D'\) for some
interior positive normalized equidistant edge weighting \(\omega\)
is the rooted
caterpillar depicted in Figure~\ref{figure:lower:bound:upgma}(a).
This is also the tree constructed by UPGMA on input~\(D'\).

From \(D'\) we construct the dissimilarity
\(D\) by putting \(D(x,y) = D'(x,y)\), for all
\(\{x,y\} \in \left ( \binom{X}{2} - \{\{1,n\}\} \right )\),
and \(D(1,n) = D(n,1) = s\) for some constant \(s > 2\).
Note that UPGMA will still generate the tree \(T\) on
input~\(D\) and that
\(\sigma_D(T) \geq \frac{s}{2(n-1)}\) holds.

Now consider any binary rooted phylogenetic tree \(T'\) on \(X\) whose
structure is as depicted in Figure~\ref{figure:lower:bound:upgma}(b).
More specifically, the rooted subtree \(T_1\) has 
\(\lfloor \frac{n}{2} \rfloor\) leaves, one of them
being \(1\), and \(T_2\) has 
\(\lceil \frac{n}{2} \rceil\) leaves, one of them
being \(n\). Using Lemma~\ref{lemma:basic:properties:NEME:score}(i) 
and the fact that $D'(x,y)\leq 2$ for all \(\{x,y\} \in \binom{X}{2}\), we have
\[
\sigma_D(T')
\leq \alpha^{T'}_{\{1,n\}} \cdot D(1,n)  + \sum_{\{x,y\} \in \binom{X}{2}} \alpha^{T'}_{\{x,y\}} \cdot D'(x,y)
\leq \frac{2s}{n^2 - 1} + (n-1),
\]
implying that \(\frac{n}{4} \sigma_D(T') \leq \frac{s}{2(n-1)} \leq \sigma_D(T)\)
for \(s \geq \frac{1}{2}n(n+1)(n-1)^2\).
But this implies
\(\sigma_D(T) \geq \frac{n}{4} \min \{ \sigma_D(T') : T' \in {\bf R}_X\}\),
as required.
\epf

Next we briefly touch upon another potential approach from the literature
for approximating the minimum NEME score.
To describe this approach, note that the NEME problem is related to
the problem of finding a \emph{sparsest cut}, that is, 
given a dissimilarity \(D\) on a finite set \(X\)
compute a \emph{split} \(A|B\) of \(X\), that is, a bipartition
of \(X\) into two non-empty subsets \(A\) and \(B\), such that
\[
\frac{1}{|A| \cdot |B|} \sum_{a \in A,\, b \in B} D(a,b)
\]  
is minimum. This problem is usually stated in terms
of edge weighted graphs and known to be NP-hard.
Recent work on this problem mainly
concentrated in finding good approximations
of a sparsest cut (see e.g. \cite{aro-haz-10a}).

Interestingly, a greedy top-down analogy of UPGMA
based on recursively splitting \(X\) by sparsest
cuts has been proposed for detecting hierarchical 
community structure of social networks \cite{man-mat-08a}.
Using again the dissimilarity \(D\) described in the proof of 
Lemma~\ref{lemma:least:squares:lower:bound},
it can be seen, however, that this approach can yield
trees whose NEME score is worse by a quadratic factor in \(|X|\)
than the minimum NEME score.

In view of the fact that the approaches we explored
so far have not led to good approximation guaranties,
we apply in the following a generic approach from
the literature to establish a polylogarithmic
approximation guaranty for dissimilarities that
satisfy the triangle inequality, that is, \emph{metrics}. 
This approach relies on two ingredients: 
\begin{itemize}
\item[(i)]
The existence of
a polynomial time algorithm with polylogarithmic 
approximation guaranty for treelike metrics.
\item[(ii)]
The fact \cite{fak-rao-04a}
that there exists a polynomial time algorithm 
that computes, for any metric \(D\) on a set \(X\)
with \(n\) elements, a collection \(D_1,D_2,\dots,D_k\)
of treelike metrics on \(X\) along with positive
coefficients \(\beta_1,\beta_2,\dots,\beta_k\), 
\(\beta_1 + \beta_2 + \dots + \beta_k = 1\), such that
\begin{itemize}
\item[(1)]
\(D(x,y) \leq D_i(x,y)\) for all \(i \in \{1,2,\dots,k\}\)
and all \(x,y \in X\), and
\item[(2)]
there exists a constant \(c > 0\) such that
\[
\sum_{i=1}^{k} \beta_i D_i(x,y) \leq c \cdot \log_2(n) \cdot D(x,y)
\]
for all \(x,y \in X\).
\end{itemize}
\end{itemize} 

We shall first establish~(i). To this end, we rely on the
following fact that, phrased in various guises, seems to be
mathematical folklore. For the convenience of the reader we
provide a short proof and phrase it in terms
of splits in \emph{unrooted} binary phylogenetic 
trees, that is, trees obtained from 
rooted binary phylogenetic trees by removing the root and the edge incident 
with it, and then suppressing the resulting degree two vertex. 
Recall that each edge $e$ in an unrooted binary phylogenetic tree $T$ on $X$ 
induces a split $S_e=A_e|B_e$ of \(X\) in which $A_e$ and $B_e$ are 
the leaf sets of the two connected components resulting from removing 
$e$ from $T$. 

\begin{lemma}
\label{lem:centre:unrooted}
In every unrooted binary phylogenetic tree \(T\) on $X$ with $|X| \geq 2$
there exists an edge $e$ such that the split \(A_e|B_e\) of \(X\) satisfies 
\begin{equation}
\label{equation:balanced:tree:partition}
\frac{1}{3}n 
\leq \min \{ |A_e|, |B_e|\}
\leq \max \{ |A_e|, |B_e|\}
\leq \frac{2}{3}n.
\end{equation}
\end{lemma}

\pf
Assume that such an edge does not exist. Replace all edges
\(e\) of \(T\) by a directed edge in such a way that this 
directed edge points to the larger of the two sets \(A_e\) and \(B_e\). Then every
directed edge incident with a leaf of \(T\) is directed away from that
leaf and, in view of the fact that all other vertices of \(T\)
have degree three, one of those vertices must be such that
all three directed edges incident with this vertex \(v\) are
directed towards it. But this implies that, while all edges \(e\) 
incident with \(v\) must clearly satisfy \(\min \{ |A_e|, |B_e|\} \leq \frac{n}{2}\),
by the pigeonhole principle, 
at least one of these edges must also satisfy \(\min \{ |A_e|, |B_e|\} \geq \frac{n}{3}\),
contradicting our assumption.
\epf

Note that Lemma~\ref{lem:centre:unrooted} does not hold for
non-binary trees. The next lemma establishes~(i). Recall that a 
\emph{treelike metric} on \(X\) is a metric for which there exists
an unrooted phylogenetic tree \(T'\) on \(X\)
with a non-negative edge-weighting \(\omega'\)
with \(D=D_{(T',\omega')}\).

\begin{lemma}
\label{lemma:approximation:factor:treelike:metrics}
Let \(D\) be a treelike metric on a set \(X\) with
\(n\) elements. Then a binary rooted phylogenetic tree
\(T\) on \(X\) with 
\[
\sigma_D(T) \leq 
c \cdot \log_2(n) \cdot \min \{ \sigma_D(T') : T' \in {\bf BR}_X\}
\]
for some positive constant \(c\) can be computed in
 time $O(n^2)$.
\end{lemma}

\pf
Let \(T'=(V',E')\) be a binary unrooted phylogenetic tree on \(X\)
and \(\omega'\) a non-negative edge weighting of \(T'\)
with \(D=D_{(T',\omega')}\). For every edge \(e \in E'\)
we denote by \(D_e\) the metric that assigns~1 to
a pair \((x,y) \in X \times X\) if the path from
\(x\) to \(y\) in \(T\) contains edge \(e\). Otherwise
\(D_e\) assigns the value~0 to \((x,y)\). Then 
$D= \sum_{e \in E'} \omega'(e) \cdot D_e(x,y)$
and hence by Equation~(\ref{equation:score:linear}) 
we have,
for any rooted phylogenetic tree \(T\) on \(X\),
\begin{align*}
\sigma_D(T) 
&= \sum_{e \in E'} \omega'(e) \sigma_{D_e}(T)
\geq \sum_{e \in E'} \frac{\omega'(e)}{2},
\end{align*}
where the last inequality follows from the fact
that, for all \(e \in E'\), \(D_e\) is an ultrametric and,
therefore, \(\min \{ \sigma_{D_e}(T'') : T'' \in {\bf BR}_X\} = \frac{1}{2}\)
by Theorem~\ref{theorem:consistency:NEME:binary:trees}.

Hence, it suffices to show how to construct
in polynomial time a rooted phylogenetic tree \(T\) on \(X\) with 
\(\sigma_{D}(T) \leq 
     c \cdot \log_2(n) \cdot \sum_{e \in E'} \frac{\omega'(e)}{2}\)
for some constant \(c > 0\). This is done recursively as follows.
Using Lemma~\ref{lem:centre:unrooted}, we find an edge \(e \in E'\) such that the split 
\(S_e = A_e|B_e\) of \(X\) induced by \(e\) satisfies (\ref{equation:balanced:tree:partition}).
We require that the resulting
rooted phylogenetic tree \(T\) on \(X\) will have
the clusters \(A_e\) and \(B_e\). Then we remove \(e\)
from \(T'\). This yields, after suppressing the two vertices
of degree~2, two unrooted phylogenetic trees \(T'_{A_e}\) on \(A_e\) 
and \(T'_{B_e}\) on \(B_e\) which represent the restriction of
\(D\) to \(A_e\) and \(B_e\), respectively. If \(|A_e| > 1\) (\(|B_e| > 1\))
we construct a binary rooted phylogenetic tree \(T_{A_e}\) on \(A_e\)
(\(T_{B_e}\) on \(B_e\)) recursively. Otherwise \(T_{A_e}\) (\(T_{B_e}\)) 
is the unique rooted phylogenetic tree on \(A_e\)  (\(B_e\)).
Then we glue the roots
of \(T_{A_e}\) and \(T_{B_e}\) together and add a new root~\(\rho\)
to obtain a binary rooted phylogenetic tree \(T=(V,E,\rho)\) on \(X\).
Note that this construction can clearly be done in time $O(n^2)$
and, since we choose the edges for recursively partitioning \(T'\) in such a way that 
Inequalities~(\ref{equation:balanced:tree:partition}) hold,
it follows that every path in $T$ contains \(O(\log_2(n))\) vertices. 

\begin{figure}
\centering
\includegraphics[scale=1.0]{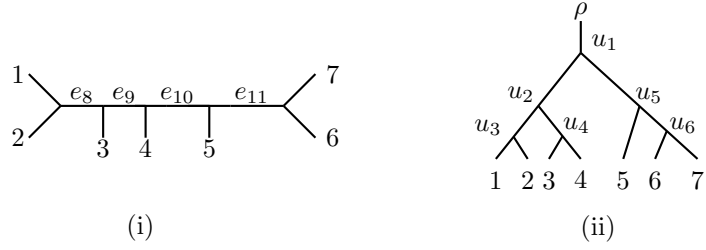}
\caption{An illustration of the canonical map \(\varphi\) used in the 
proof of Lemma~\ref{lemma:approximation:factor:treelike:metrics}: 
(i) An unrooted binary phylogenetic tree $T'$ on \(X=\{1,2,\cdots,7\}\) (for simplicity, the label $e_i$ 
of the edge incident with leaf $i$ is omitted).  
(ii) A binary rooted phylogenetic tree on \(X\) constructed from $T'$ by consecutively 
removing the following sets of edges 
\(\{e_{10}\},\{e_8\},\{e_5,e_{11}\},\{e_3,e_4,e_9\},\{e_6,e_7\},\{e_1,e_2\}\). Then the canonical map 
\(\varphi\) is as follows: \(\varphi(e_{10})=u_1, \varphi(e_{8})=u_2, \varphi(e_5)=\varphi(e_{11})=u_5, 
\varphi(e_3)=\varphi(e_4)=\varphi(e_9)=u_4, \varphi(e_6)=\varphi(e_7)=u_6,\) 
and \(\varphi(e_{1})=\varphi(e_{2})=u_3\).}
\label{figure:canonical:map} 
\end{figure}

Note that the construction of \(T\) induces the canonical map
\(\varphi:E' \rightarrow (V - (X \cup \{\rho\}))\)
that assigns to \(e \in E'\) the internal vertex
\(v\) of \(T\) that was constructed as the root in a recursive step
when edge \(e\) was removed from \(T'\) (see Fig.~\ref{figure:canonical:map} for an example).  
Also note that \(\varphi\) is, by construction,
surjective. It is, however,
not injective because suppressing degree~2 vertices
leads to clusters of original edges in \(T'\) that 
are removed together at a single recursive step.

Next, for each edge \(e \in E'\) let  \(\omega_e=\omega_{(D_e,T)}\) and 
let  $\Gamma_e$ be the set of those vertices $v \in (V - (X \cup \{\rho\}))$ 
for which there exist \(x,y \in X\) 
such that (i) \(v=lca(x,y)\) and (ii) \(e\) lies on the path from \(x\) to \(y\)
in \(T'\). By construction of $T$, the set $\Gamma_e$ consists precisely of
those vertices in \(V - (X \cup \{\rho\})\) that lie on the path from
\(\varphi(e)\) to \(\rho\) in \(T\), implying that \(|\Gamma_e| \in O(\log_2(n))\).
In addition, Equation~(\ref{equation:optimal:edge:weights}) 
implies that $D_e$ contributes at most \(1/2\)
to \(h_{(T,\omega_e)}(v)\) for all $v \in \Gamma_e$ and,
for all \(v \in V - \Gamma_e\) we have \(h_{(T,\omega_e)}(v)=0\). Therefore, 
using Equation~(\ref{equation:total:length:from:heights}) to obtain
the second equality below, we have 
\begin{eqnarray}
\label{equation:upper:bound:score}
\sigma_D(T) 
&=& \sum_{e \in E'} \omega'(e) \sigma_{D_e}(T) \notag\\
&=& \sum_{e \in E'} \omega'(e) \Big(\sum_{v \in (V - (X \cup \{\rho\}))} h_{(T,\omega_e)} (v) \Big) \notag\\ 
&=&\sum_{e\in E'} \omega'(e) \Big( \sum_{v\in \Gamma_e}  h_{(T,\omega_e)}(v)  \Big) \notag\\
&\leq& c \cdot \log_2(n) \cdot \sum_{e \in E'} \frac{\omega'(e)}{2}
\end{eqnarray}
for some constant \(c > 0\), as required.
\epf

\begin{figure}
\centering
\includegraphics[scale=1.0]{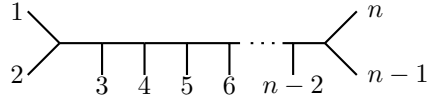}
\caption{An unrooted caterpillar tree on \(X=\{1,2,\dots,n\}\).
         The metric \(D\) induced by this tree on \(X\) (all edges
         have weight~1) is such that a rooted phylogenetic tree
         on \(X\) with minimum NEME score for \(D\) cannot
         by obtained by rooting the caterpillar somewhere.}
\label{figure:rooting:caterpillar} 
\end{figure}

Note that there are treelike metrics \(D\) on \(X\)
for which a binary rooted phylogenetic tree with
minimum NEME score cannot be obtained by rooting
the unrooted tree representing \(D\) somewhere.
That means that the structure of the unrooted 
tree representing \(D\) need not reflect much the
structure of the rooted trees with minimum NEME score
for \(D\). Consider, for example, the metric
\(D\) on \(X=\{1,2,\dots,n\}\), \(n \geq 6\), induced by the 
unrooted caterpillar in Figure~\ref{figure:rooting:caterpillar}.
All edges are assigned weight~1. Then the 
recursive algorithm in the proof of 
Lemma~\ref{lemma:approximation:factor:treelike:metrics}
yields a binary rooted phylogenetic tree \(T\) on \(X\)
with \(\sigma_D(T) \in O(n \log_2 n)\) by
the upper bound in~(\ref{equation:upper:bound:score}). In contrast,
for any binary rooted phylogenetic tree \(T'\) obtained
by rooting the caterpillar, there must exist, for all
\(k \in \{3,4,\dots,\lfloor \frac{n}{2} \rfloor\}\), at
least one vertex \(v\) in \(T'\) with \(|ch(v)| = k\) and
\(h_{(T',\omega)}(v) \geq \frac{1}{2(k-1)}(k+(k-1)+\dots+4+3) \geq \frac{k}{4}\), where
\(\omega=\omega_{(D,T')}\). This implies that
\(\sigma_D(T')  \geq \sum_{k=3}^{\lfloor \frac{n}{2} \rfloor} \frac{k}{4} \in \Omega(n^2)\).
The next theorem summarizes our results on
approximating the NEME score for metrics.

\begin{theorem}
\label{theorem:approximation:metric}
Let \(D\) be a metric on a finite set \(X\) with
\(n\) elements. Then a rooted binary phylogenetic tree
\(T\) on \(X\) with 
\[
\sigma_D(T) \leq 
c \cdot \log^2(n) \cdot \min \{ \sigma_D(T') : T' \in {\bf BR}_X\}
\]
for some positive constant \(c\) can be computed in
polynomial time.
\end{theorem}

\pf
Let \(D_1,D_2,\dots,D_k\) be a collection of treelike
metrics for \(D\) together with 
coefficients \(\beta_1,\beta_2,\dots,\beta_k\)
as described in~(ii) above. In addition, let
\(T^*\) be a binary rooted phylogenetic tree on \(X\)
with minimum NEME score for \(D\) and, similarly,
\(T_i^*\) be a binary rooted phylogenetic tree on \(X\)
with minimum NEME score for \(D_i\), \(1 \leq i \leq k\). 
Assume that \(\sigma_{D_1}(T_1^*) \leq \sigma_{D_i}(T_i^*)\)
for all \(i \in \{1,2,\dots,k\}\). Finally, let
\(T_1^{o}\) be the binary rooted phylogenetic tree on \(X\)
constructed for \(D_1\) using the recursive algorithm
in the proof of Lemma~\ref{lemma:approximation:factor:treelike:metrics}.
Then we have, using repeatedly the linearity from Equation~(\ref{equation:score:linear}):
\begin{align*}
\sigma_D(T_1^{o})
&\leq \sigma_{D_1}(T_1^{o})\\
&\leq c' \cdot \log_2(n) \cdot \sigma_{D_1}(T_1^*)\\
&\leq c' \cdot \log_2(n) \cdot \sum_{i=1}^k \beta_i \cdot \sigma_{D_i}(T_i^*)\\
&\leq c' \cdot \log_2(n) \cdot \sum_{i=1}^k \beta_i \cdot \sigma_{D_i}(T^*)\\
&= c' \cdot \log_2(n) \cdot \sigma_{(\sum_{i=1}^k \beta_i \cdot D_i)} (T^*)\\
&\leq c'c'' \cdot \log_2^2(n) \cdot \sigma_D(T^*),
\end{align*}
where \(c'\) and \(c''\) are positive constants that
come from the upper bound (\ref{equation:upper:bound:score}) and
property (2) of the collection \(D_1,D_2,\dots,D_k\), respectively.
Hence \(\sigma_D(T_1^{o}) \leq c \cdot \log_2^2(n) \cdot \sigma_D(T^*)\)
for some constant \(c>0\) 
and the tree \(T_1^{o}\) can be constructed in polynomial time.
\epf

Interestingly, the above approach can be applied to 
\emph{any} variant of minimum evolution as long as the objective
function is a linear combination of the input distances and
the variant is consistent (the latter trivially implies
ingredient~(i) above {\tw and, thus, saves a
factor of $\log n$ in the approximation guaranty}). In particular, the original unrooted 
ME problem \cite{rzh-nei-93a} has these properties and can, therefore,
be approximated for metrics {\tw within a factor of \(O(\log_2 n)\)}.
To the best of our knowledge, this is the first non-trivial 
approximation result for the unrooted ME problem.

\section{Concluding remarks and open problems}
\label{section:conclusion}

In this paper, we have highlighted some properties of the UPGMA
method. We now conclude by pointing out two possible directions for future work.
The first direction concerns improving the approximation guarantee
for the NEME problem presented in the last section. 
Recall that the interpretation
of the balanced minimum evolution score of an unrooted
tree as an average over spanning cycles has been used
(as one ingredient amongst others) in \cite{fio-jor-12a} 
to design a constant-factor polynomial
time approximation algorithm for the balanced
minimum evolution problem in case
the given dissimilarity satisfies the triangle
inequality. We expect that the results
presented in this paper can similarly serve as the basis for
a better understanding of the approximation properties
of the NEME problem. A concrete conjecture we have in this 
direction is that UPGMA always generates a tree whose NEME score
is within a factor in \(O(n)\) of the minimum score.

The second direction for future work concerns the
so-called safety radius of the NEME approach for
computing rooted phylogenetic trees. The 
safety radius concept was introduced to quantify
how much distortion of the input distance matrix a method
can tolerate and still return the correct tree
(see e.g.~\cite{atteson1999performance}). For example, 
it is known that in the rooted-tree setting 
UPGMA has a safety radius of 1 \cite{gas-mck-04a}, 
and that both 
neighbor joining and BME-based tree construction have a safety radius of $\frac{1}{2}$ 
(see~\cite{atteson1999performance} and \cite{par-gui-10a}, respectively).
We conjecture that the safety radius of NEME-based tree construction 
tends to~0 as
the number of leaves of the trees tends to infinity.
In this context, it might also be of interest to study the 
stochastic safety radius of the NEME problem, a concept that
was recently introduced~\cite{gascuel2015stochastic}, and
which aims to understand consistency within a probabilistic setting.


\end{document}